\documentclass[12pt,preprint]{aastex}
\newcommand{\pref}{\protect\ref}
\newcommand{\solrad}{\ifmmode{R}_{\rm S}\else${R}_{\rm S}$\fi}
\newcommand{\solmas}{\ifmmode{M}_{\rm S}\else${M}_{\rm S}$\fi}

\newcommand{\ctn}{\ifmmode\kappa\else$\kappa$\fi}

\newcommand{\flxu}{$\,$ergs$\,$cm$^{-2}\,$s$^{-1}$}

\newcommand{\term}[2]{\mbox{$\,^{#1}{\rm #2}$}}
\def\term#1 #2/{\mbox{$\,^{#1}{\rm #2}$}}

\def\aspcs{{ASP Conf.\ Ser.}}

\renewcommand{\vec}[1]{{\bf #1}}
\newcommand{\cross}{\times}

\newcommand{\jcb}{\ifmmode\vec{j}\cross\vec{B}\else$\vec{j}\cross\vec{B}$ \fi}
\newcommand{\coords}[2]{$x,y$=#1\arcsec,#2\arcsec}
\newcommand{\coordm}[2]{$x,y$=#1,#2\,Mm}

\newcommand\tabone{
\protect\begin{deluxetable}{llll}
\renewcommand{\baselinestretch}{1.0}
\tablecaption{Sensitivity of MDI, KPVT and Hinode-SP
  longitudinal magnetograms \label{tab:sens}}
\tablehead{Instrument/mode & noise per pixel & pixel size & noise in
  flux \\
 & Mx~cm$^{-2}$ & arc seconds & units of 10$^{15}$ Mx  }
\startdata
MDI/full disk & 17 & $1\farcs 984\times 1\farcs984$ & 350\\
KPVT/synoptic & 2.8 & $1\farcs 148\times 1\farcs148$& 19\\
Hinode SP/normal map & 3& $0\farcs 164\times 0\farcs164$& 0.42\\
(Kitt Peak 40 channel magnetograph & 0.4 &$\dag$ & $\approx13$ \\
\citealp{Livingston+Harvey1971}) & & &\\
\enddata 
\tablecomments{1\arcsec{} on the Sun corresponds to 725 km
  \protect\citep{
Allen1973}. $^\dag$Seeing limited, here we use an effective pixel size
  of $2.5\times2.5\arcsec$ corresponding to half of the quoted resolution of
  $5\arcsec$. }
\end{deluxetable}

}

\newcommand\figone{
\begin{figure}[!ht] 
\epsscale{0.9}
\plotone{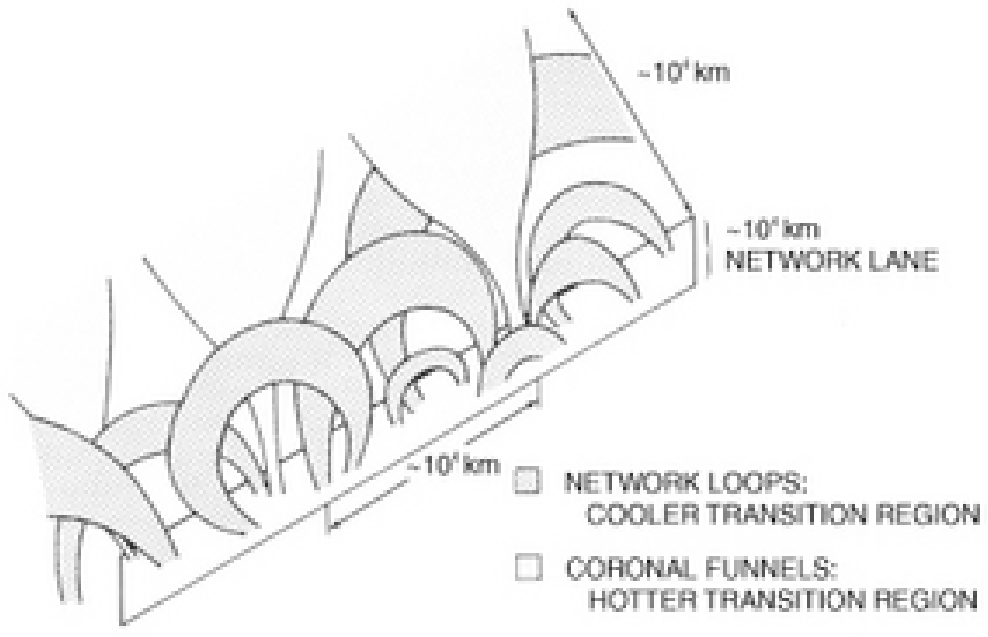}  
\caption{\label{fig:dowdy86} A sketch of the likely magnetic structure in
  a quiet region of the Sun's magnetic network, according to 
  \protect\citet{Dowdy+Rabin+Moore1986}.  The ``coronal funnels''
  are similar  to the structures modeled by \protect\citet{Gabriel1976},
  which can account for emission above about $2\times10^5$ K.  The
  ``network loops'', arising from mixed polarity magnetic fields
  {\em within} network boundaries, explain the cool transition region
  emission, in Dowdy's picture.}
\end{figure}
}

\newcommand\figtwo{
\begin{figure}[!ht] 
\epsscale{1.15}
\plotone{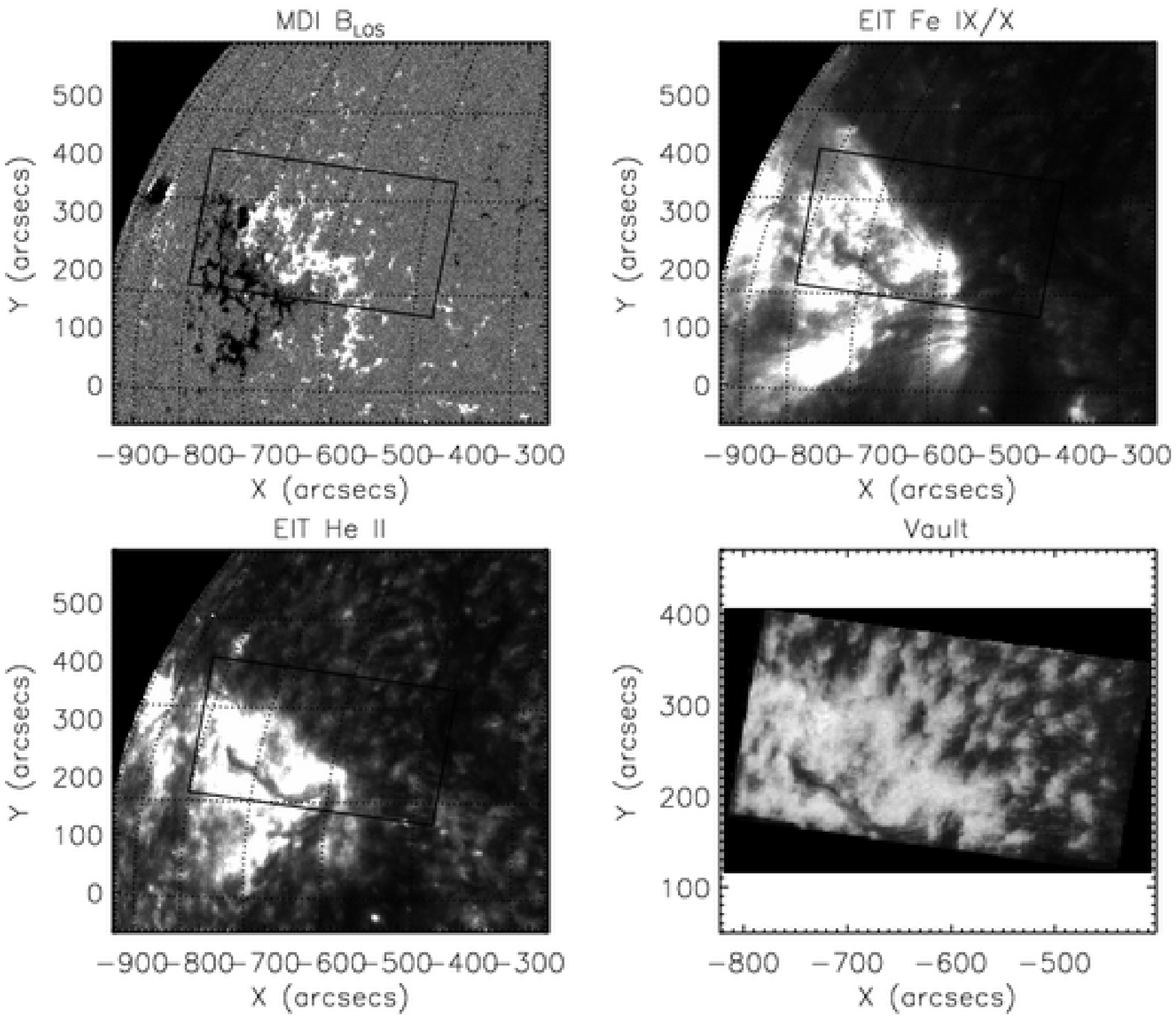}  
\caption{\label{fig:context} Context images showing the VAULT FOV
  (boxed region) for the observation at 18:12 UT on 14 June
  2002. Shown are 
   images of longitudinal field from MDI (between $\pm 200$ Mx~cm$^{-2}$), and of
  He II 304 emission and 171 \AA{} Fe IX/X emission from EIT, both
  instruments on the SOHO spacecraft.  These images were
  differentially rotated to 18:12:00 UT,
  the epoch of these VAULT-2 observations.
  The active region in the SE
  corner contains a filament along a magnetic neutral line, another filament
  is seen in the He II image, oriented roughly N-S on the western edge
  of the VAULT FOV. Also shown are (lower right) VAULT-2 data. 
}
\end{figure}
}

\newcommand\figfour{
\begin{figure}[!ht] 
\epsscale{.98}
\plotone{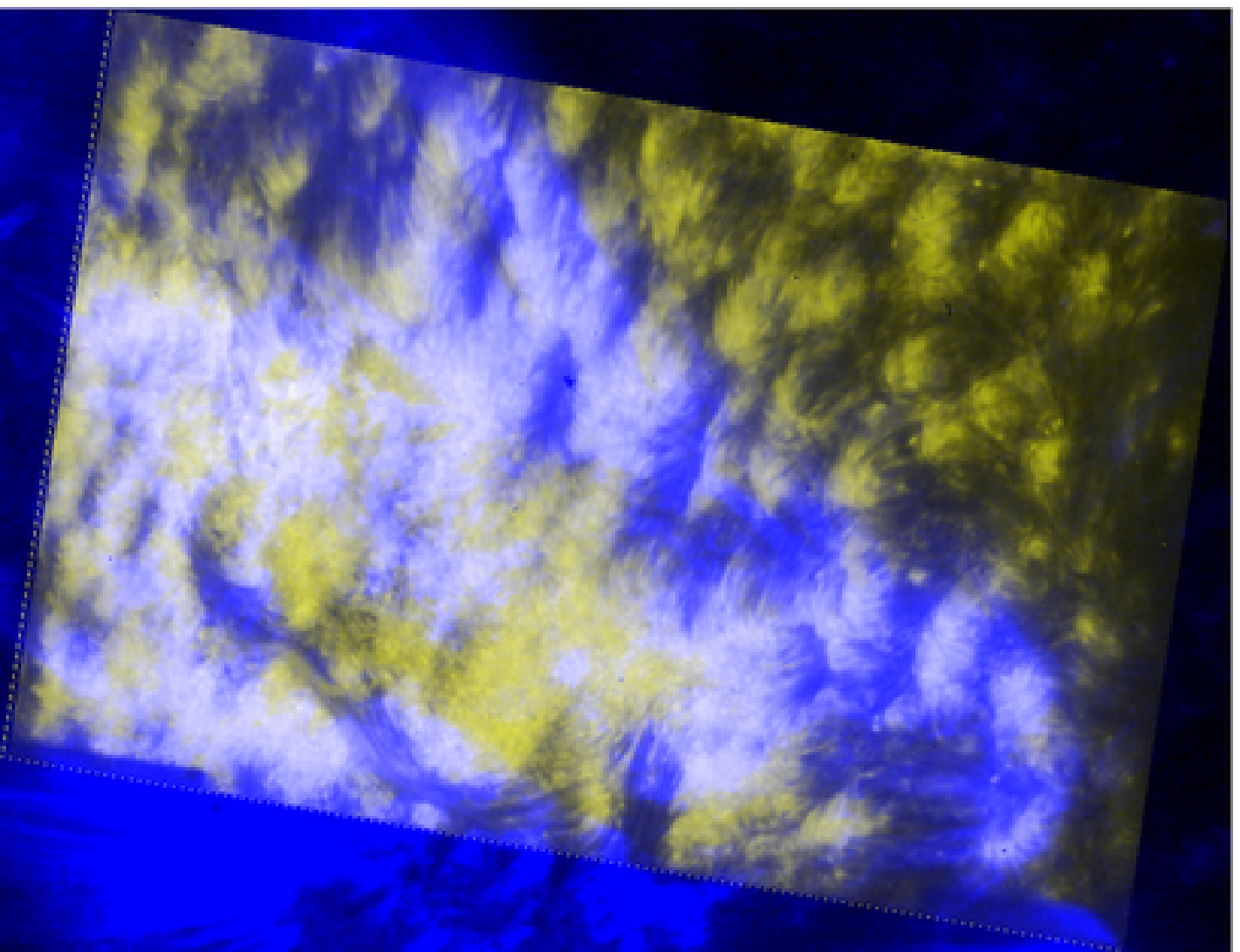}  
\caption{\label{fig:tricolor} A false color image of 
VAULT-2 data
(yellow) together with TRACE 171\AA{} data of the lower corona
(blue).  The TRACE data were 
obtained 30 minutes before the VAULT-2 image.  Four TRACE 171\AA{} images
were added together.
}
\end{figure}
}

\newcommand\figfive{
\begin{figure}[!ht] 
\epsscale{.6}
\includegraphics[angle=90,scale=0.45]{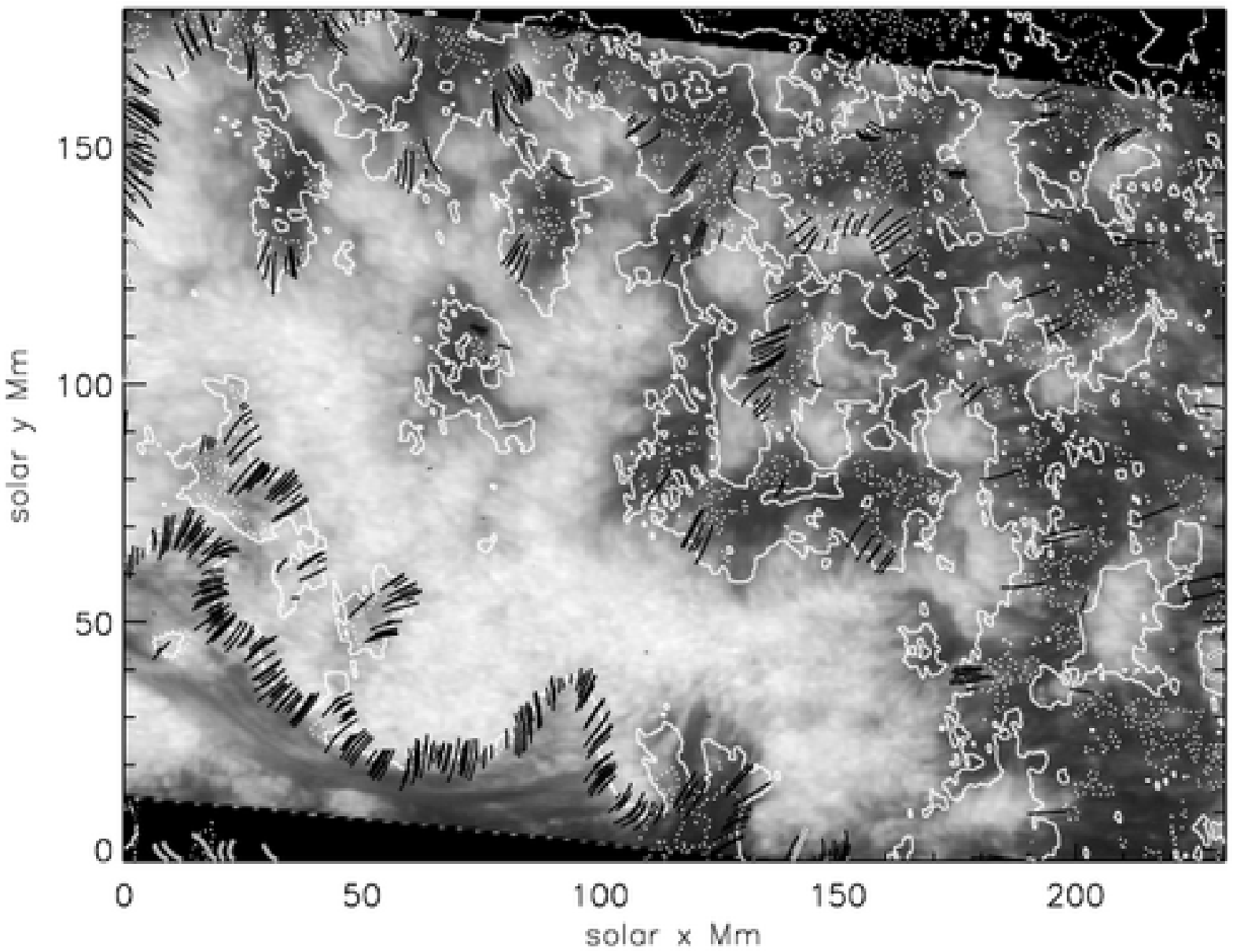}
\includegraphics[angle=90,scale=0.45]{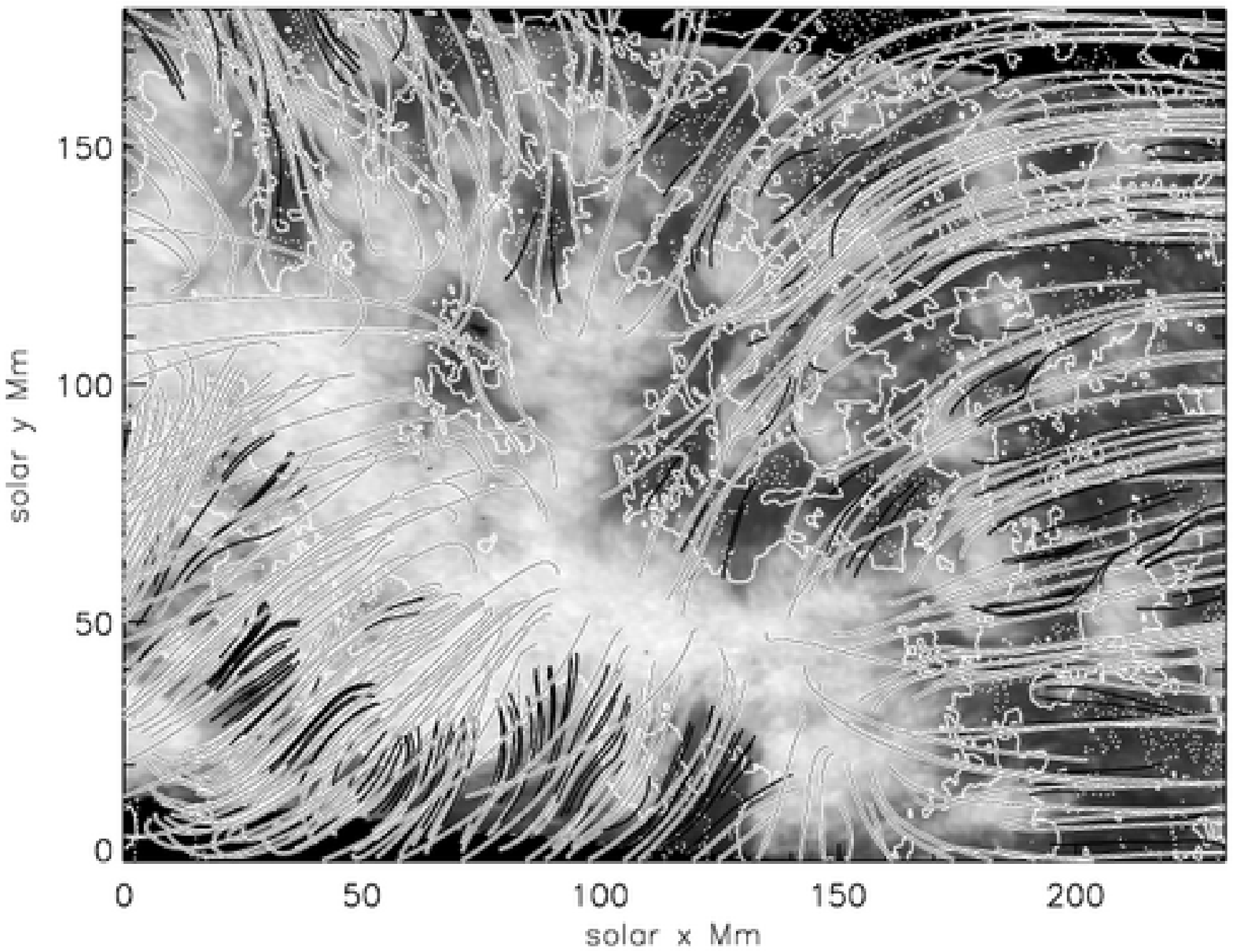}
\caption{\label{fig:kpvtvault} Magnetic field contours and 
potential field lines superposed on the VAULT-2 image. The upper panel
shows loops with  lengths $<$10 Mm, the lower shows 
longer loops.  Contours are at 
$\pm 2\sigma$ 
($\pm 5.6$ Mx cm$^{-2}$),  negative
contours are dashed lines, positive solid.  Not all field lines from
each pixel are plotted even if their signal exceeds the noise (see
text), to avoid confusion. Field lines reaching heights 
$\le 5$ Mm are plotted as a black line, others are
shown as a black on top of a white line. 
The figure origins are the same but
arbitrary, the center of the figure is Sun center.
}
\end{figure}
}

\newcommand\figfivec{
\begin{figure}[!ht] 
\epsscale{1}
\includegraphics[width=0.55\textwidth,clip,angle=90]{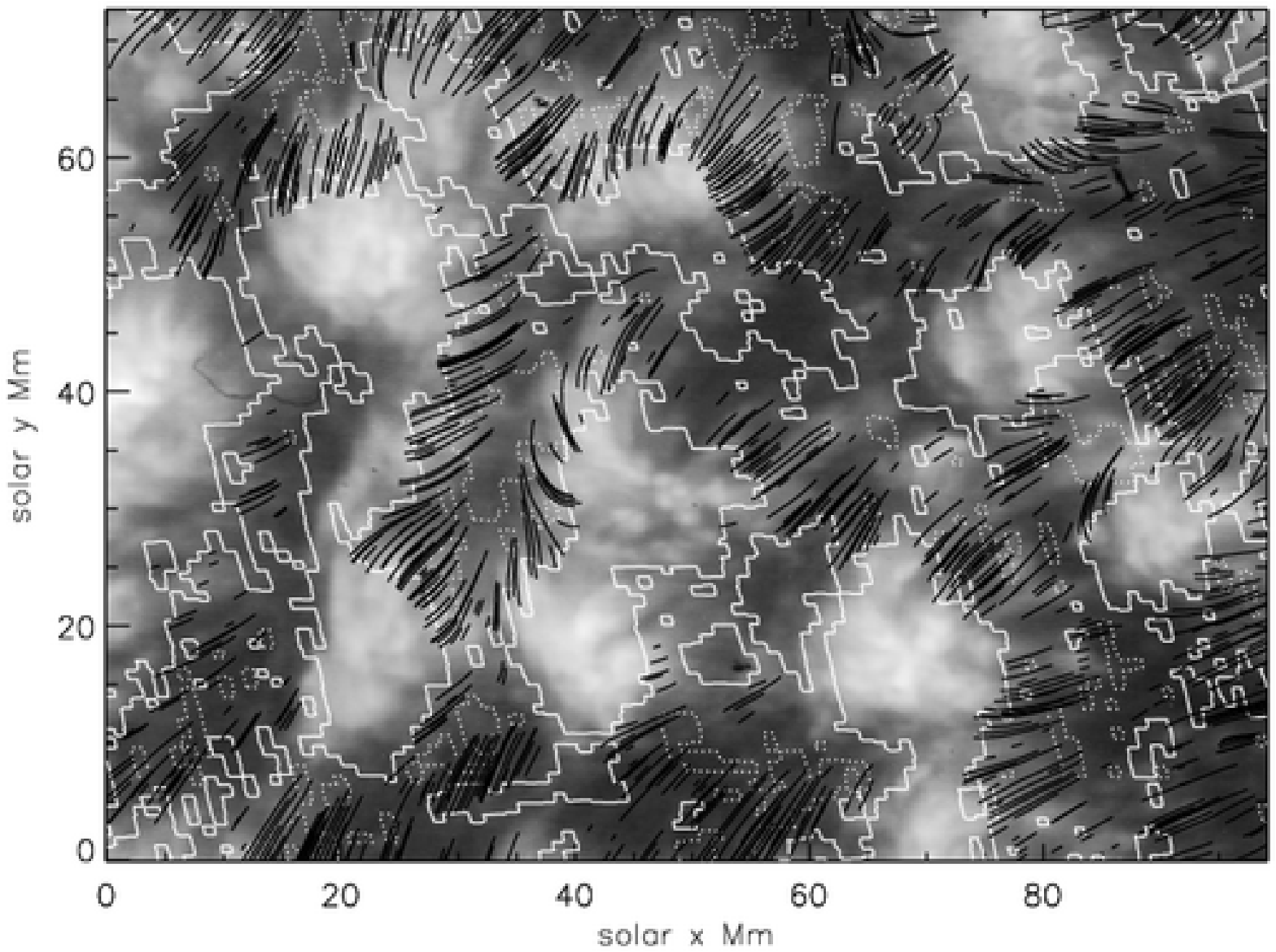}
\includegraphics[width=0.55\textwidth,clip,angle=90]{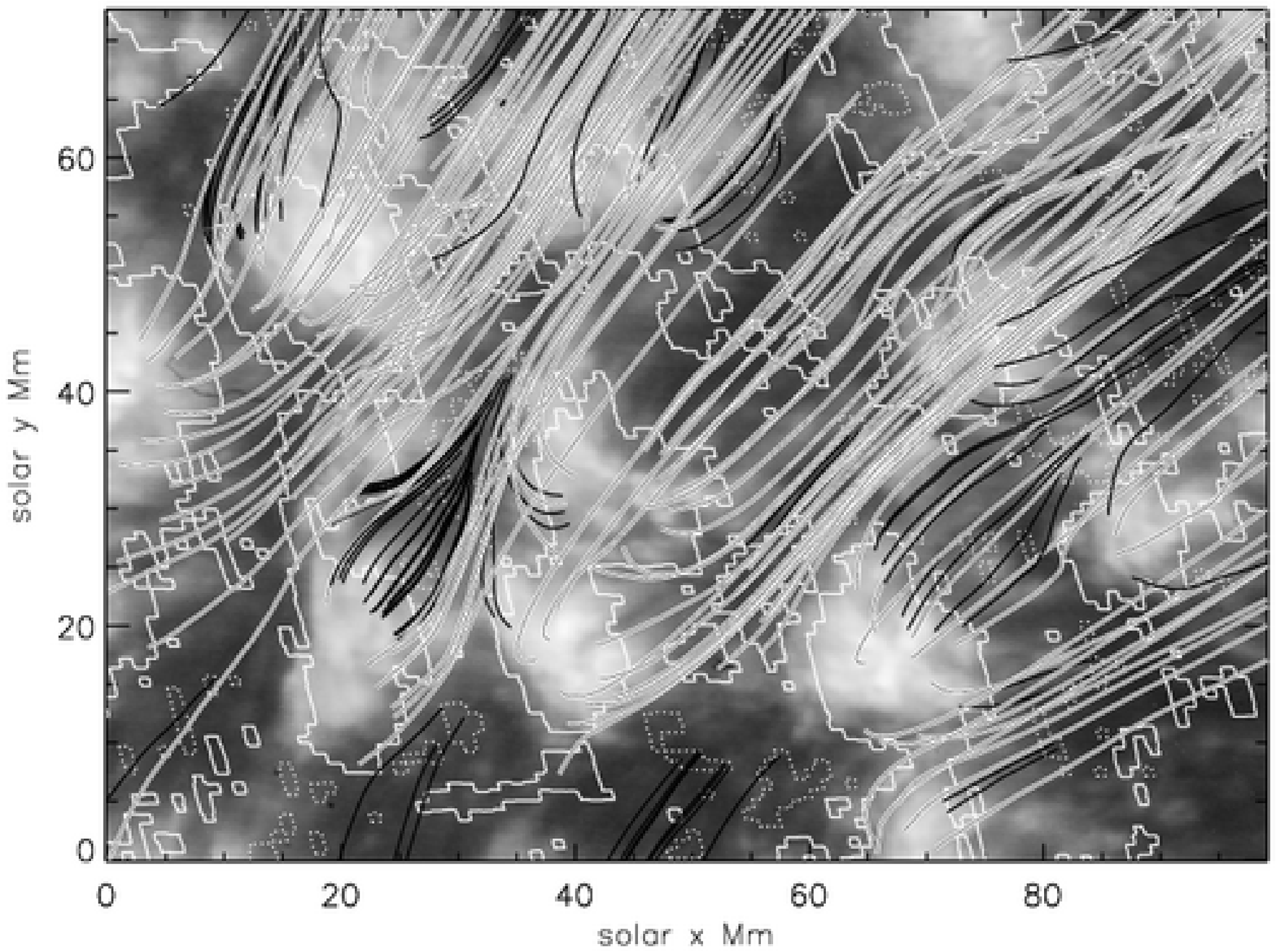}
\caption{\label{fig:kpvtvaultsm} 
Potential field lines 
superposed on a portion of the VAULT-2 
image, centered at 
\coords{-500}{+290} in Figure \pref{fig:context}.  The area covers
perhaps 10 supergranular areas.   The upper panel shows loops of 10Mm
length and less, the lower panel longer loops. Contours of $\pm
2\sigma$ 
($\pm 5.6$ 
Mx cm$^{-2}$) are shown, where $\sigma$ is the rms uncertainty in
line of sight field strength.  Field lines were plotted with no limit
set on the signal-to-noise ratios of the magnetogram. 
There is little correspondence between 
short loops and the bright L$\alpha$ emission, instead the brightest
emission originates from the bases of loops longer than 10Mm. 
}
\end{figure}
}

\newcommand\figsix{
\begin{figure}[!ht] 
\epsscale{1}
\includegraphics[width=0.6\textwidth,clip,angle=90]{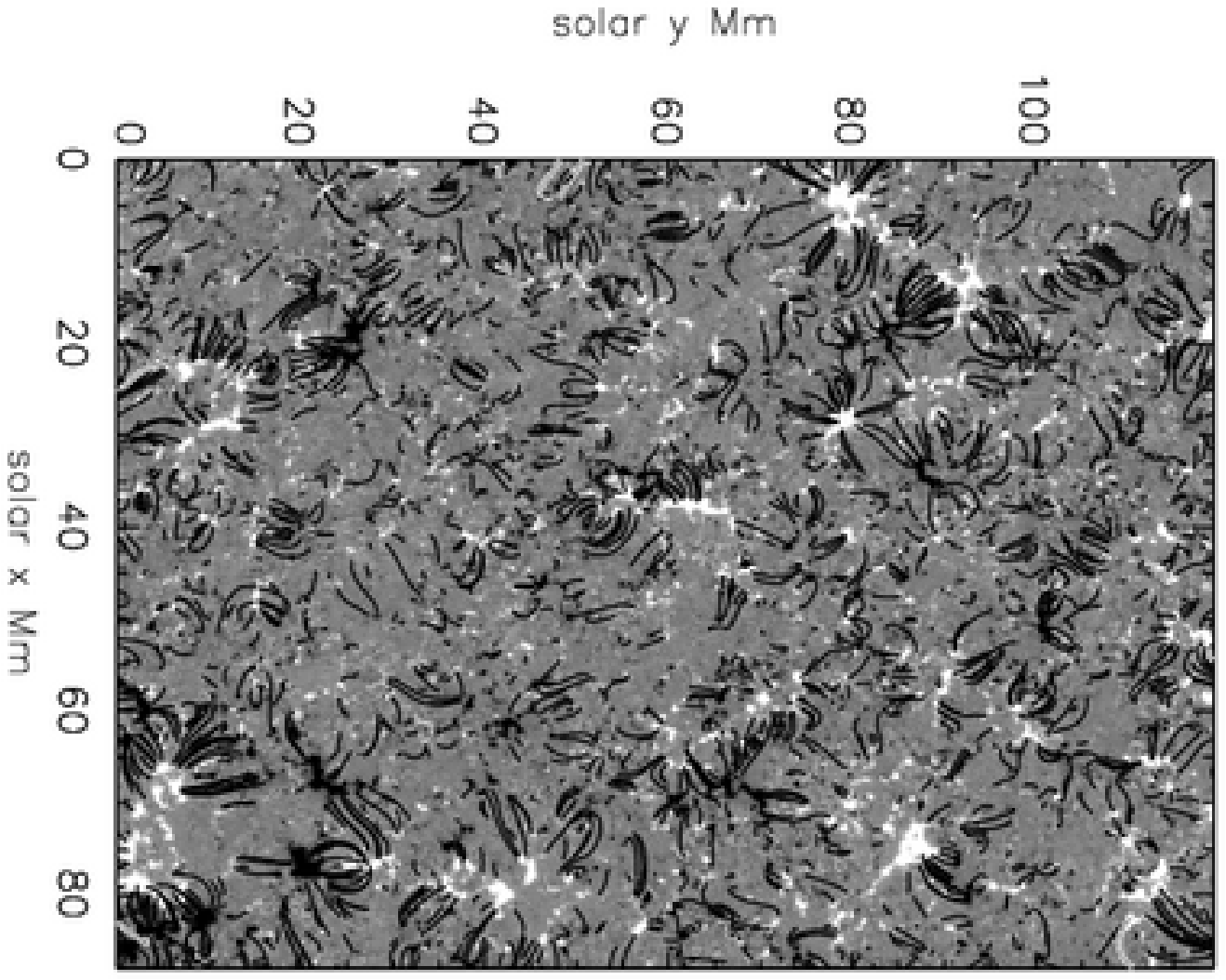}
\includegraphics[width=0.6\textwidth,clip,angle=90]{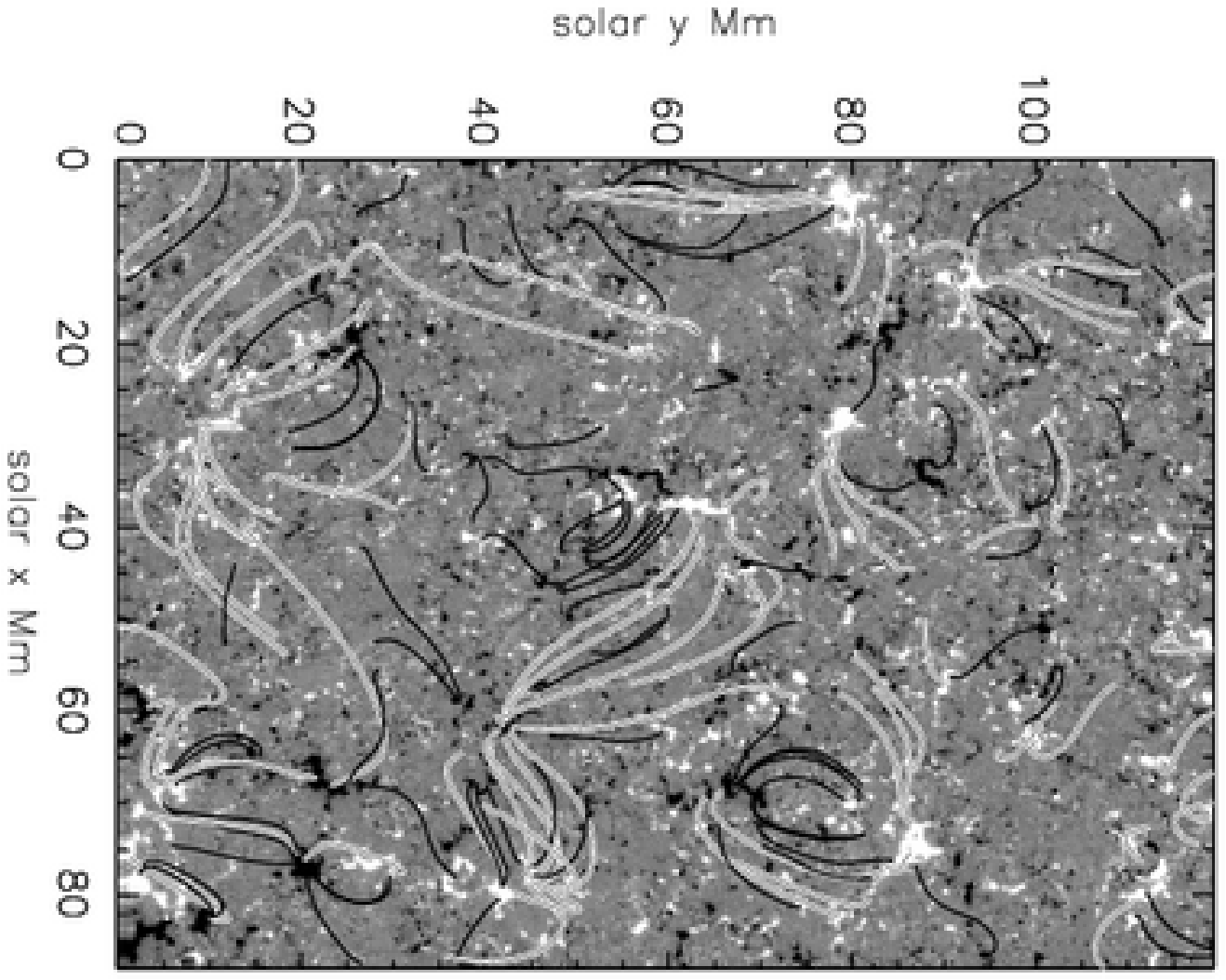}
\caption{\label{fig:spqr} 
Line of sight field strength images,
obtained with the SP on Hinode on 17 November 2007 beginning at 11:31 UT,
 are shown with 
potential field lines superposed.  
The magnetograms are shown on a linear scale between 
-50 and +50 Mx~cm$^{-2}$ to show weak flux regions.
The upper panel
shows loops with total length below 10 Mm, the lower shows those with
longer lengths. The coordinates have an
arbitrary origin.
}
\end{figure}
}

\newcommand\figseven{
\begin{figure}[!ht] 
\epsscale{.98}
\includegraphics[width=0.6\textwidth,clip,angle=90]{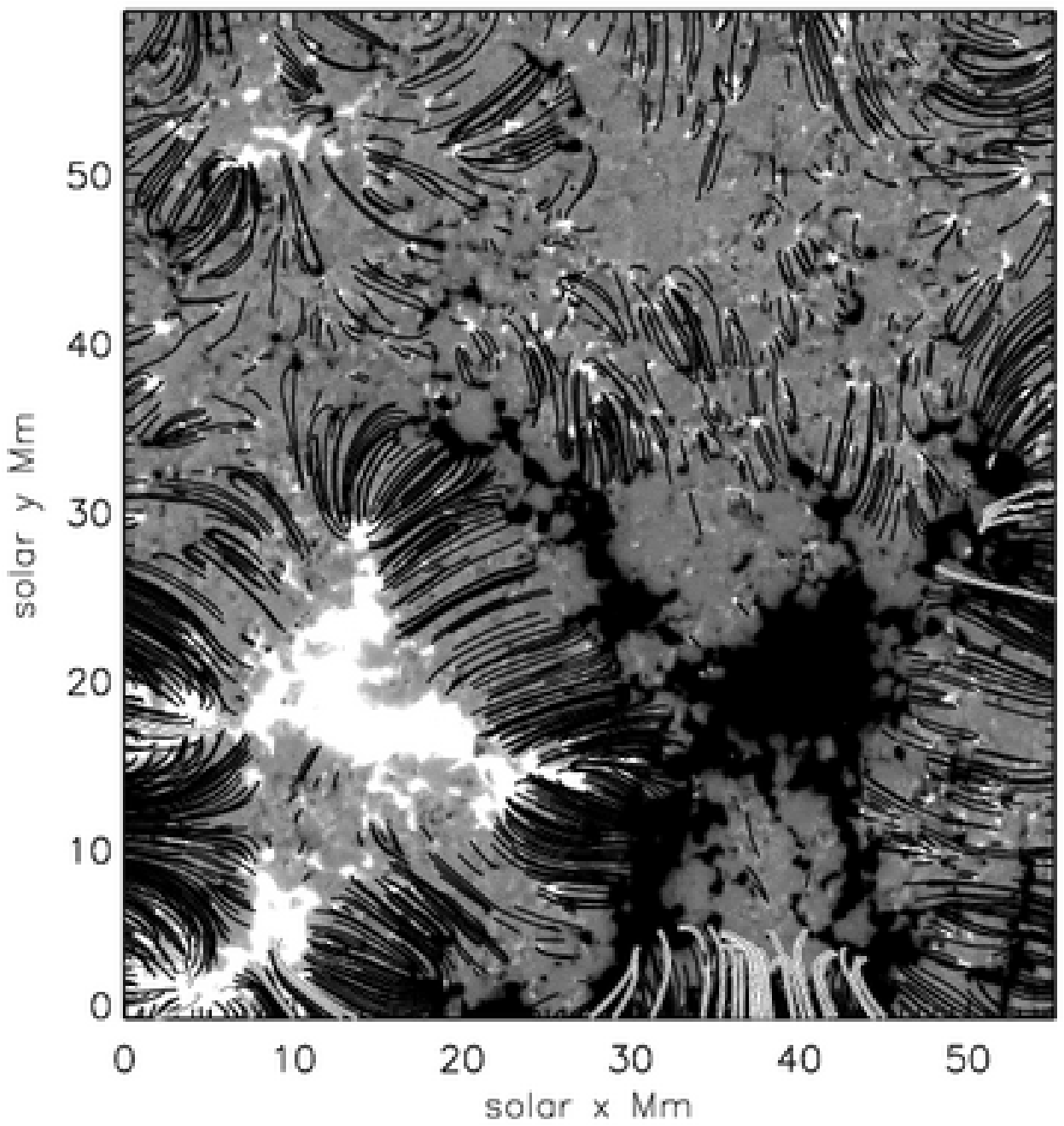}
\includegraphics[width=0.6\textwidth,clip,angle=90]{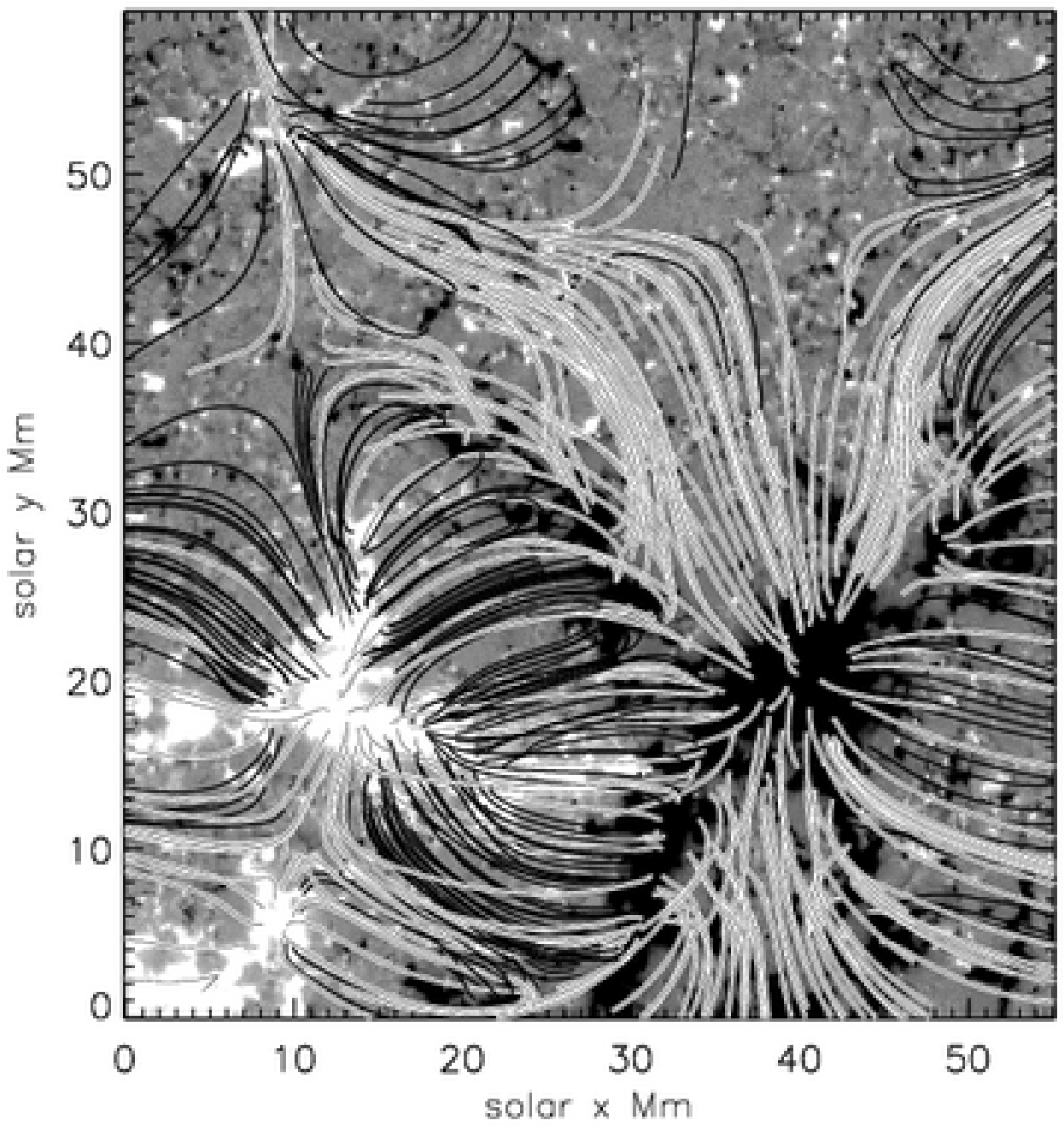}
\caption{\label{fig:spar} A similar plot to 
figure \pref{fig:spqr}, but for the  active region observed with the SP
on Hinode, on 29
December  2006 beginning at 01:51 UT.  (Field line cusps near $y=47$ are artifact of
the Fourier method used.) 
}
\end{figure}
}

\newcommand\figeight{
\begin{figure}[!ht] 
\epsscale{.98}
\includegraphics[width=0.6\textwidth,clip,angle=90]{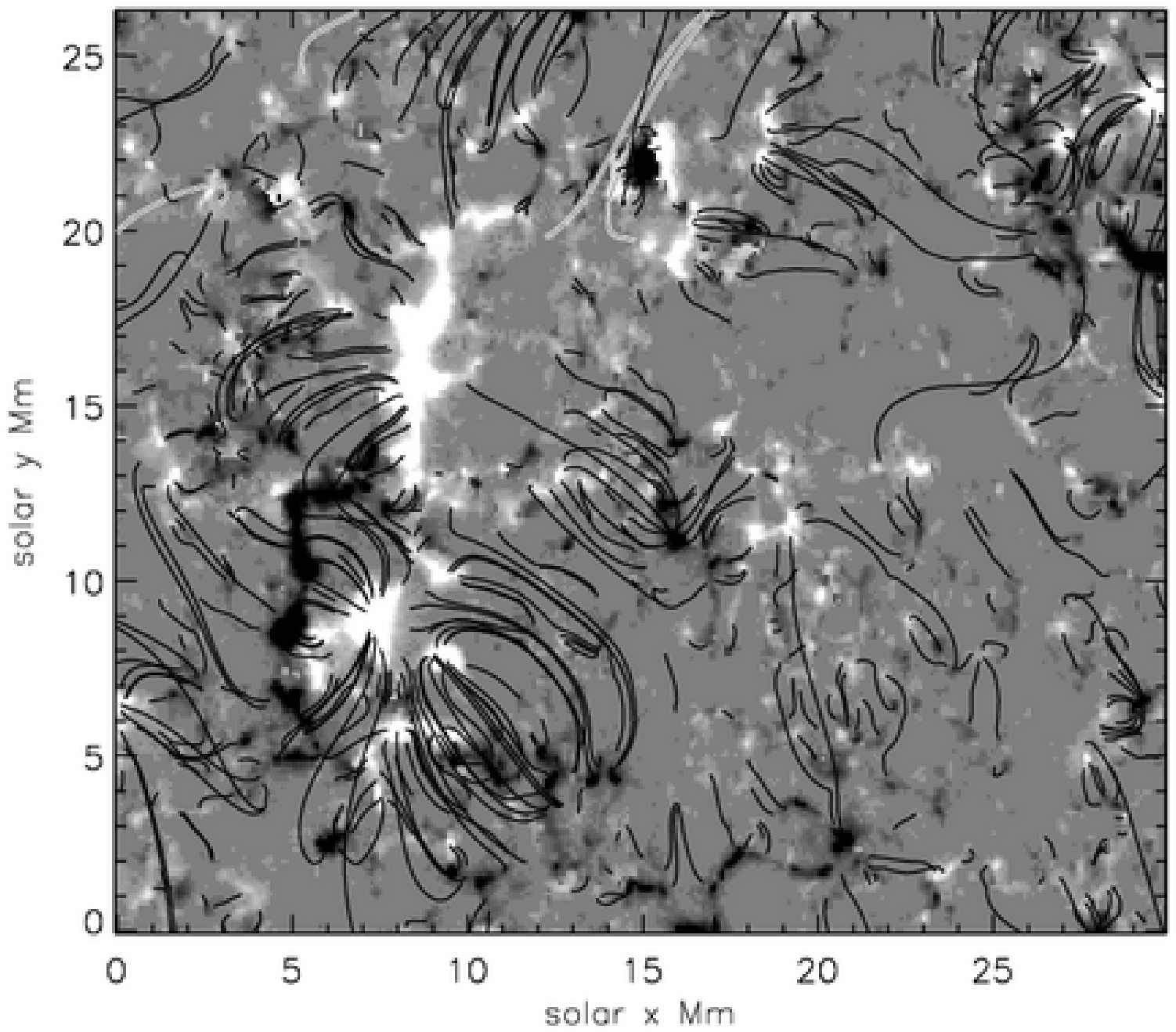}
\includegraphics[width=0.6\textwidth,clip,angle=90]{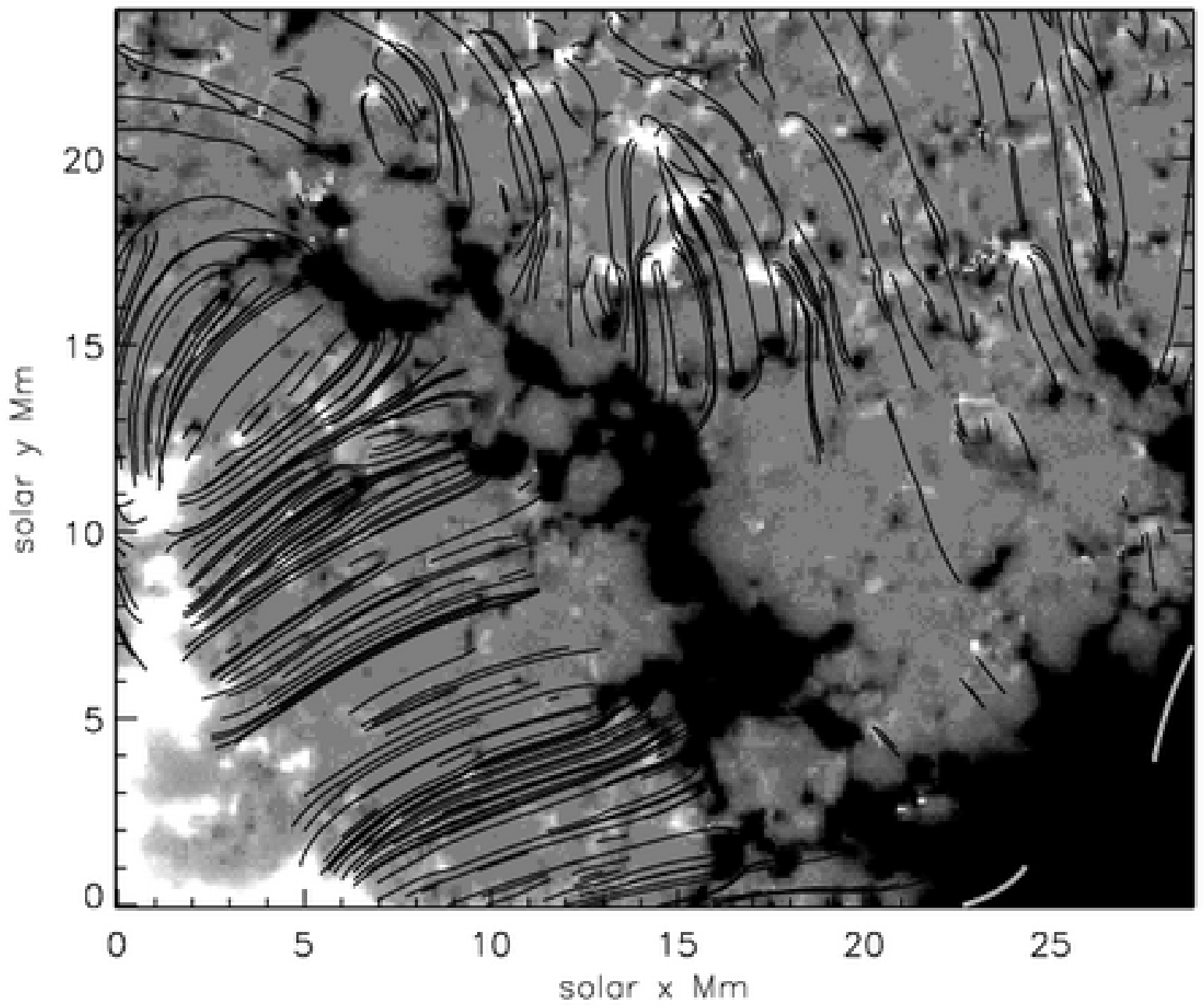}
\caption{\label{fig:spqrc} Close-up magnetograms of flux
concentrations observed by the SP on Hinode 
from the central portions of the quiet region dataset from 
November 27 2007 (upper panel),
and the more active area from 
and 29 December 2006 (lower panel). 
The regions show cover about the area of one supergranule. 
Short magnetic loops abound in both regions, as the strong network
concentrations return to cell interiors of opposite polarity or to
nearby concentrations. 
}
\end{figure}
}

\newcommand\fignine{
\begin{figure}[!ht] 
\epsscale{.8}
\plotone{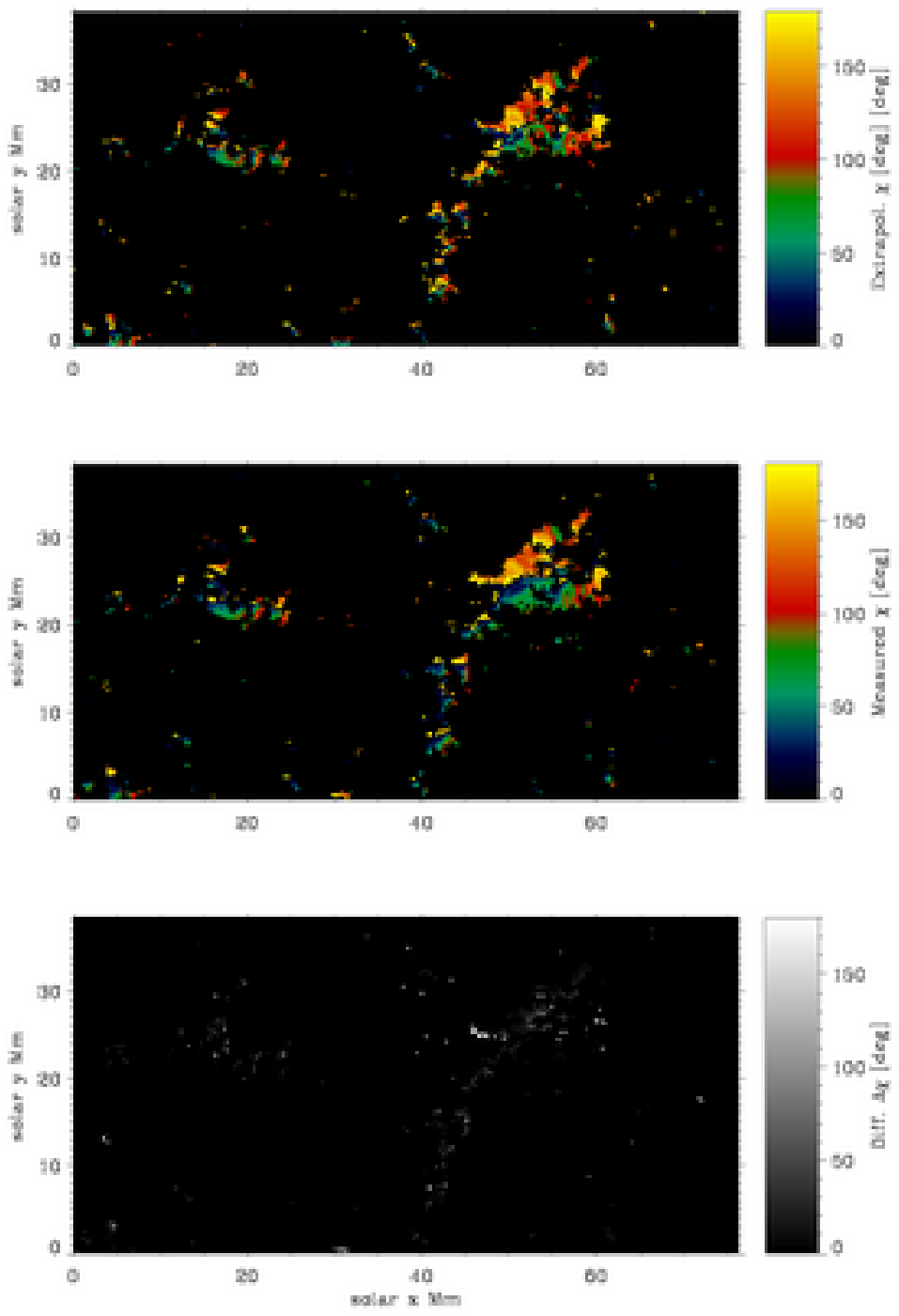}
\caption{\label{fig:azimuths} Magnetic field azimuths from inversions 
  of the measured 
  Hinode SP data (top), those computed from the potential field approximation
  using just the line-of-sight component of the magnetic field
  (center), and the difference (bottom).  
  The plot shows the lower half of the dataset obtained on 29
  December 2006 which contains significant Stokes $Q,U$
  signals. White regions in the bottom panel are most likely the
  result of the 180$^\circ$ ambiguity in inverted fields. The qualitative agreement  suggests that
  such differences, if typical,  do not affect our conclusions concerning
  the locations of loops of various lengths. 
}
\end{figure}
}

\slugcomment{}

\begin{document}

\title{\large On the magnetic structure of the solar transition region}

\author{Philip Judge and Rebecca Centeno}
\affil{High Altitude Observatory,
National Center for Atmospheric Research\altaffilmark{1},
P.O. Box 3000, Boulder CO~80307-3000, USA\\ \vbox{}}

\begin{abstract}

We examine the hypothesis that ``cool loops'' dominate emission
from solar transition region plasma below temperatures
of $2\times10^5$K.  We compare published VAULT 
images of H L$\alpha$, a lower transition region line, 
with near-contemporaneous magnetograms from Kitt Peak, obtained during
the second flight (VAULT-2) on 14 June 2002.
The measured surface fields and potential 
extrapolations suggest that there are
too few short loops, and that  L$\alpha$ emission is associated with
the base regions of longer, coronal loops. 
VAULT-2 data of 
network boundaries have an
asymmetry on scales larger than supergranules, also 
indicating an association with long loops.
We complement the Kitt Peak data with very sensitive vector
polarimetric data from the Spectro-Polarimeter on board Hinode,
to determine the influence of very small magnetic
concentrations on our analysis.  From these data
two
classes of behavior are found: within the cores of strong 
magnetic flux concentrations ($> 5\times10^{18}$ Mx) associated with active network
and plage, small-scale mixed fields are absent and any short loops can
connect just the peripheries of the flux to cell interiors. Core fields
return to the surface via longer, most likely coronal, loops.  In weaker
concentrations, short loops can connect between concentrations and
produce mixed fields within 
network boundaries as suggested by Dowdy
and colleagues. The VAULT-2 data which we examined are associated with strong
concentrations.  We conclude that the cool loop model applies only to
a small fraction of the VAULT-2 emission, but we cannot discount a
significant role for cool loops in quieter regions.  We suggest a 
physical picture for how network L$\alpha$ emission may occur through
the cross-field diffusion of neutral atoms from chromospheric into coronal plasma. 

\end{abstract}


\section{Introduction}
\label{sec:introduction}

In spite of a century or so of research, the solar atmosphere
continues to challenge  our understanding.  As well as the well-known problems
of identifying the causes of coronal and chromospheric heating, other 
phenomena such as spicules, flares, explosive events, and atmospheric
dynamics in general, remain only
partly understood.  

A particularly stubborn puzzle relates to the structure of the solar
transition region (henceforth ``TR'') - plasma between the chromosphere and corona.  The
chromosphere and TR are particularly bright over
magnetic field concentrations. Outside of sunspots, the concentrations
form the supergranular network boundaries (henceforth, ``NB''s).  Cell
interiors (``CI''s) are less bright.  The supergranular network
pattern eventually disappears at coronal temperatures $T_e \ga 10^6$ K
\citep{Tousey1971,Reeves1976}.  The upper TR (where the
electron temperature $T_e$ lies between say $2\times 10^5$ K and
coronal temperatures of $10^6$ K) is adequately described by magnetic
field-aligned thermal conduction down from the corona
\citep{Gabriel1976,Jordan1980b}, but that the lower TR
($10^4$ K $< T_e < 2\times 10^5$ K) is not so easily understood
\citep[see, for example, the reviews
by][]{Mariska1992,Anderson-Huang1998}.  Models dominated by heat
conduction along magnetic field lines fail to produce enough bright
emission from the lower transition by orders of magnitude. In other
words, the differential emission measure in the lower transition
region is far higher than predicted by such models. 
Two points argue,
however, in favor of conduction as a source of energy for the lower
TR: First, the conductive flux is of the right order of
magnitude to explain the radiation losses there. Second, the relative
intensities of coronal and TR lines vary surprisingly 
little 
over the solar surface (i.e., the differential emission
measure has a universal shape), suggesting an energetic link.  

Attempts to fix the problem have either largely
failed, or raised questions of an equally troubling nature.  For
example, models based on heat flow parallel to magnetic field lines
were explored which dropped earlier restrictions on magnetic field
geometry and a static picture. But such calculations fail to account
for the brightness of the lower transition
region (\citealp{Gabriel1976}, \citealp{Pneuman+Kopp1978}, \citealp{Athay1981},
\citealp{Woods1986}).  A variation on such models was
developed by
\citet{Cally1990}, introducing enhanced 
heat fluxes using transport by turbulent eddies.  Cally found
that the mixing length (a free parameter) should scale as
$T_e^{-1.5}$ to account for observations. But this is somewhat
unsatisfactory because there is no  physical reason why the
turbulent transport should behave like this.
\citet{Ashbourn+Woods2001} presented a more promising 
model in which the ion-acoustic
instability sets in owing to the high electron drift speed associated
with heat conduction, in the
middle TR.   The instability makes the
perpendicular ion thermal conductivity large and dominant below 10$^5$
K, thereby providing a lower temperature gradient and higher emission
measure than fluid approximations give.  
The downward heat flux was also 
explored by \nocite{Fontenla+Avrett+Loeser1990,Fontenla+Avrett+Loeser1991,
Fontenla+Avrett+Loeser1993,Fontenla+Avrett+Loeser2002} 
Fontenla {\em et al.} (1990,1991,1993,2002; henceforth ``FAL'') in
multi-fluid 
models where diffusion and bulk flows transport cool material to
hot regions, thereby extracting and radiating much of the available
energy in the corona. 

The essential property of all such models is that they produce a
geometrically thin lower TR, where emission is confined to a layer just a few tens of
km deep.  As discussed in a series of papers by Feldman and
colleagues \citep[e.g.,][]{Feldman1983}, UV, EUV and X-ray images 
present serious challenges for such thin models.   Such observations
have prompted others to set aside conduction, instead focusing on
local sources of heat within the TR, such as Joule
dissipation (e.g., \citealp{Rabin+Moore1984},
\citealp{Roumeliotis1991}) or parameterized forms of heat
dissipation \citep[e.g.][]{Antiochos+Noci1986,
Patsourakos+Gouttebroze+Vourlidas2007}.  Such models are similar to
chromospheric models \citep[e.g.][]{Vernazza+Avrett+Loeser1981},
in the sense that local heating is balanced by radiation
losses\footnote{However, unlike the chromosphere, at transition
region temperatures the plasma is almost fully ionized and there is no
internal heat sink (latent heat of ionization) which acts, in the
chromosphere, as a thermostat.  Locally heated and cooled models of
the TR are therefore susceptible to thermal instability
\citep{Cally+Robb1991}.}.  In such models, conduction plays no role,
and instead the 
heights of the structures or ``cool loops'' 
are physically limited by the pressure scale height, the
temperature scale heights being larger
\citep{Antiochos+Noci1986}.  Even in loops near $10^5$K, the height
cannot exceed 5Mm for equilibrium to exist, for heating functions
which are not of a special form \citep{Cally+Robb1991}.  Limb
observations of active regions reveal cool material much higher, but
such structures are almost always dynamic
\citep[e.g.][]{diGiorno+other2003}.

\citet{Dowdy+Rabin+Moore1986} pointed out that one would expect
 magnetic field of mixed polarity {\em within} the supergranular NBs, 
a feature absent from the model of \citep{Gabriel1976}.
Dowdy and colleagues suggested TR emission 
below $\sim 10^5$ K is dominated by radiation from locally heated loops
which connect such opposite polarity photospheric fields,  on
scales of $\la 10$Mm.   Their picture is illustrated by
figure~\pref{fig:dowdy86}, taken from their 1986 article.

\citet{Sanchezalmeida+others2007} searched for 
magnetic signatures of footpoints of cool loops, using data from
the SUMER \citep{Wilhelm+others1995a} 
and MDI \citep{Scherrer+others1995} 
instruments on the SOHO spacecraft, and G band bright point data
from the Dutch Open Telescope \citep{Hammerschlag+Bettonvil1998}. 
Their analysis was inconclusive, 
because, as we will find, neither SUMER nor MDI (full-disk) data 
have sufficient spatial resolution ($1\farcs5$ and $4\arcsec$
respectively),  or magnetic sensitivity.  

The present paper studies the proposal of Dowdy et al. by examining
very high angular resolution ($\approx 0\farcs3$) images of the lower
transition from the VAULT instrument 
\citep{Korendyke+others2001}, and comparing these with
magnetic fields and extrapolations from nearly-simultaneous
magnetogram data.  Since the magnetic data available do not have the
sensitivity of recent data, we also examine spectroheliograms of the
Sun from the spectropolarimeter 
(SP, \citealp{Lites+Elmore+Streander2001}) on the Hinode spacecraft 
\citep{Kosugi+others2007}.   We will argue that several observed
properties of the VAULT L$\alpha$ emission are not compatible with the cool loop
hypothesis.

\section{UV data of the cool transition region and corona}

We have re-analyzed data described by
\citet{Patsourakos+Gouttebroze+Vourlidas2007}, augmented by 
UV and EUV data from SOHO and TRACE \citep{Handy+others1999}. The VAULT data
were obtained during the second instrument flight 
(VAULT-2) on 14 June 2002 near 18:12 UT.  The field of view is
$375\arcsec \times 257\arcsec$ centered near coordinates
($-600\arcsec,+260\arcsec$) relative to solar disk center.  
The instrument captured both active
plage areas south and east of the center of the FOV, and 
the quieter areas to the NW.
The pixel size is $0\farcs125$ and resolution $\approx 0\farcs3$.  
Figure~\pref{fig:context} shows contextual images of the photosphere
(MDI instrument) and corona (EIT, \citealp{Delaboudiniere+others1995}) 
from the SOHO spacecraft, together with a VAULT-2
image\footnote{The image analyzed is {\tt img04.png} from {\tt
http://wwwsolar.nrl.navy.mil/rockets/vault/archives.html}.}.  
The MDI, EIT 171\AA{} and EIT 304\AA{} images were 
acquired at 17:59:30, 18:04:37, and 17:45:15 UT respectively.
The VAULT-2 image was coaligned 
with the the EIT He~II 304\AA{} image, crudely by eye. It was found to be
centered near $(-612\arcsec,260\arcsec)$, and rotated clockwise by
$9^\circ$. Furthermore the pixel sizes were found to be $0\farcs125$
and $0\farcs115$ in $x$ and $y$ respectively.  (The different pixel
scales found are not in disagreement with measurements by the VAULT
team, 
C. Korendyke, private communication 2008). 
This co-alignment is
reliable only to within $\approx \pm 2-3$\arcsec{}, but is sufficiently
accurate for our purposes. 

Focusing on the ``quiet'' NW sector of the images, the brightest
L$\alpha$ emission occurs over fairly strong, unipolar magnetic flux
concentrations (see section \pref{sec:magvault}).  A feature of the
data receiving much attention
\citep[see][in particular their fig.~2]{Patsourakos+Gouttebroze+Vourlidas2007} is the fine
thread-like structure.  The threads are most obvious seen against the
darker CIs in the neighborhood of the bright concentrations.  Similar
fine structure has been seen for many years in the related, but
notoriously difficult to interpret, H$\alpha$ line
\cite[e..g.][]{Kiepenheuer1953,Martin+others1994}.  The chromospheric
network in L$\alpha$ contains many such threads, combining to form a
collective network emission of order 10$\arcsec$ width \citep[][their
fig.~2]{Patsourakos+Gouttebroze+Vourlidas2007}.  These authors
showed that the fine structure is missing from data with lower
angular resolution than $\approx1\arcsec$.  In their picture, much of
the emission from each concentration originates from the threads, 
which are assumed to be small  loops.

\citet{Patsourakos+Gouttebroze+Vourlidas2007} also noted that ``The
threads are, at a first approximation, radially distributed or
slightly bent around the cell centers, which suggests that they could
correspond to closed structures, i.e. small loops with temperatures in
the temperature formation range of Ly$\alpha$.''
However, such patterns are not actually
common in the VAULT-2 data analyzed by them. Inspection of Fig.~2 of
\citet{Patsourakos+Gouttebroze+Vourlidas2007}, an area of
$100\arcsec\times60\arcsec$ containing say 4 or so supergranular
cells, shows that the threads are all oriented between $\approx
-40^\circ$ and $10^\circ$ of the direction of the $y$ axis (see also
figure~\pref{fig:tricolor}).  Thus {\em the thread orientations are
organized on supergranular scales (i.e. much larger than granules), 
and larger still near active regions}.
The
bright L$\alpha$ patches appear like small ``comets'' whose tails are
comprised of the threads, the ``heads'' of the comets pointing
generally towards the active region in the SE of the image (bottom
left of the figure).  Such thread orientations are at odds with
magnetic loop footpoints which might be expected from more random 
granular and
supergranular forcing of surface magnetic fields.  The ``comets'' are 
reminiscent of patterns seen in H$\alpha$ data, particularly in the
neighborhood of filaments, for many years
\citep{Martin+others1994,Low1996}.

Figure~\pref{fig:tricolor} shows the VAULT-2 data (yellow) together with
TRACE 171\AA{} data (four summed images) 
of the lower corona (blue), obtained close to
17:45 UT. TRACE data were aligned to  VAULT-2 data using the EIT co-alignment.  
While 171\AA{} coronal emission is present almost
everywhere, it is brightest only in the vicinity of the plage and
filament.  White regions in the image shows areas where both features
are bright, including some ``moss'' emission (slightly granulated
structure) in the neighborhood of the filament, but also including
some loops slightly further from the filament region.  Use of a
different color table shows coronal moss emission throughout the plage,
with weak 171 emission even in the ``quiet'' region (upper right
corner).  171 \AA{} moss is associated with conductive heating from
overlying hotter loops \citep{Fletcher+dePontieu1999, Berger+others1999}. 
The coronal emission will be related to magnetic field
extrapolations below.

VAULT images 
from an earlier flight (VAULT-1) were analyzed by 
\citet{Korendyke+others2001,Vourlidas+others2001}.  These images
contain qualitatively similar features to those seen in the data
studied here, including moss and threads organized
into comet-like patterns in the region within $\approx$ 100\arcsec{} of the
filament.  The VAULT-1 DATA contain more quiet areas which contain
network concentrations hosting significant, but less ordered,
thread emission \citep[fig.~5 of][]{Korendyke+others2001}.

To summarize, 
the essential features of the VAULT-2 data under scrutiny are:

\begin{itemize}
  \item{} Much of the emission is organized into thin threads,
  originating in the network cell boundaries, most
  visible against the dark cell interiors, 
  \item{} near large-scale coronal structures and organized
  photospheric magnetic fields (active regions), the threads have
  characteristic orientations on scales in excess of supergranules, 
  \item{} within plage or large  active network boundaries, the
  emission appears to have a small scale granular structure associated
  with conductively heated coronal emission called 
``moss'' \citep{Vourlidas+others2001}. 
\end{itemize}

\section{Magnetic fields measured at the epoch of the VAULT-2 observations}
\label{sec:magvault}

MDI full-disk data have a 1$\sigma$ noise level of 17 Mx~cm$^{-2}$ per
$1\farcs 984\times 1\farcs984$ pixel, determined from
the spatial power spectrum assuming the noise is ``white''.
Fortunately, higher quality longitudinal magnetograms from the Kitt
Peak Vacuum Telescope synoptic program are available, here we analyze
a scan taken between 16:00 and 16:55 UT on June 14 2002.  These data
have pixels of $1\farcs 148\times 1\farcs148$, each with a noise of
$\approx 2.8$ Mx~cm$^{-2}$.  Table~\pref{tab:sens} lists the relative
sensitivities of these and other magnetograms.  The KPVT
instrument can detect the ubiquitous ``salt and pepper''
weak longitudinal field, with fluxes of $(65\rightarrow100)\times10^{15}$ Mx,
discovered many years ago by \citet{Livingston+Harvey1971}.  Such
small fluxes are invisible to MDI. 

We aligned the KPVT data with VAULT-2 by eye, using the unsigned
magnetic flux, we estimate the accuracy of the alignment is no worse
than 3\arcsec, probably better.  This apparently large uncertainty
arises because although there
is a strong correlation between chromospheric and TR emission and
absolute magnetic flux, the correlation is only on scales in excess of a few seconds of
arc \citep{Skumanich+Smythe+Frazier1975,Sanchezalmeida+others2007}. Variations with time and 
small scale spatial variations limit the 
co-alignment accuracy. 
We rotated both datasets to disk center assuming that L$\alpha$ and
the KPNO data arise from the same spherical surface.  This is
manifestly incorrect, given that L$\alpha$ is formed at least 2Mm
above the photosphere and has contributions from spicules extending to
10Mm or so. Thus after the rotation we re-did the co-alignment. Lastly,
we assumed that the observed fields are statistically radial, and
re-computed the line-of-sight field on the rotated surface.  (Our
comparisons of field line morphology with L$\alpha$ images 
are insensitive to the exact choice of radial vs. vertical
field, and our calculations are only potential anyway). 

We computed the potential magnetic field from this surface,
assumed to be flat, using the Fourier method\footnote{In all figures
except \pref{fig:spqr} and \pref{fig:spar}, a larger field of
view was used to compute the potential fields than is plotted, to try
to avoid periodic artifacts.}.  These
calculations crudely indicate the morphology of magnetic fields
overlying the surface fields, assuming that sources (i.e. currents) of
magnetic field are negligible except those sub-surface currents
responsible for the surface fields.  Figure \pref{fig:kpvtvault} shows
VAULT data, contours of magnetic flux density, and field lines
superposed.  The field lines were plotted only if both footpoints
exceeded twice the noise level of 2.8 Mx~cm$^{-2}$, 
but the
extrapolations themselves include all pixels.
In this way we reject 
more than 95\% of field lines which arise solely from noise.  
(Figure~\pref{fig:kpvtvaultsm} below includes all field lines 
regardless of signal-to-noise ratios, to illustrate  that rejection of
noisy data is not a critical issue). 
The plotted field lines were
evenly sampled to avoid overcrowding here and in later figures.  The
region observed by VAULT is dominated by flux of positive polarity.
Both short and long loops are aligned locally in generally the same
direction.  The large positive flux region connects to negative
polarity fields in the SW corner and outside of the plotted field of
view.  The overwhelming amount of positive flux guides essentially all
field lines away from it, in the potential fields shown.

Let us consider the extrapolated field lines as possible 
plasma loops, and first examine
those 
of length $\le 10$Mm. (Lengths and heights of potential field 
loops are on average related, but we also
discriminate between low and high lying loops in the following 
figures). Such short loops are
candidates for the ``cool loop'' model of L$\alpha$ emission.  If the
emission were dominated by such structures, we would expect to find a
correlation between bright thread emission and the position and
density of these field lines.  There are indeed places where this
is the case, near \coordm{70}{150} and \coordm{140}{110} in the figure,
for example, but in most cases the bright emission and short loop
densities are poorly correlated.  Figure \pref{fig:kpvtvaultsm} shows
a close up of flux concentrations\footnote{We refine our definition of
  ``concentration'' here to refer to an aggregation of flux of one polarity which defines
part of a NB. Typically these have an area of
$\ga 1$Mm$^2$ and a flux $\ga 10^{18}$ Mx, about three times the 
detection limit of MDI full disk data (table~\pref{tab:sens}).  
They
are closely correlated with ``the network'' (clumps) of L$\alpha$ emission.}
centered near \coords{-500}{+290} in
figure \pref{fig:context}, showing many more short loops and their
properties compared to the L$\alpha$ threads.  
In this
case we set no limits on the footpoint signal-to-noise ratios in
selecting field lines, to see if excluding noisy pixels 
might introduce an important bias. It does not, it simply reduces the
number of field lines with at least one footpoint in the CI regions.
Thus, in figure~\pref{fig:kpvtvaultsm}, one finds areas where cool
loops are indeed plausible (e.g. between \coords{15,60} and
\coords{55,60}), but there are other areas where short loops are found
without L$\alpha$ threads (most other locations in the figure). 
In any case, these short loops tend
to connect only the {\em peripheries} of flux concentrations to
CI regions, but L$\alpha$ is bright also directly over
the flux concentrations themselves.  

In fact, L$\alpha$ is brightest
where loops much {\em longer} than 10Mm originate in the photosphere
(lower panel of the figure).  This
is because there simply is not room for short loops to reach the central,
brightest and unipolar regions of the flux concentrations from neighboring
CI fields.  These longer and taller loops, if
they contain much plasma, are likely to have most of their volume
filled with hot (i.e. coronal) plasma in order that they have a
reasonable lifetime (i.e. close to
hydrostatic equilibrium, \citealp[e.g.][]{Rosner+Tucker+Vaiana1978}).
{\em Thus, L$\alpha$ emission from the NBs appears to be
dominated by processes near the footpoints of {\em long}
loops}. The longer loops not only originate in the bright cores of the
concentrations, but they also show ``comet''-like structure in broad
agreement with the L$\alpha$ thread morphology.  These field line shapes 
arise because surface field lines initially directed towards the SE are forced to avoid
the
strong same polarity flux concentration there, and so turn towards the
NW, attracted by opposite polarities outside of the FOV. 

These results present difficulties for the ``cool loop'' model,
which attempts to explain {\em most} of the network emission.
However, care is needed before arriving at stronger conclusions. 
At the sensitivity of the KPVT
data, there are simply too few small loops to account for all threaded
L$\alpha$ structures.  It may be that, as suggested by
figure~\pref{fig:dowdy86}, there are unresolved mixed polarities within
the cores of the flux concentrations, as well as in CI regions.
To address this problem, we turn to more sensitive
measurements of surface magnetic fields, at higher angular resolution, 
to see if such fields 
exist on the Sun at a level undetected by these data, and discuss the
role of fields unresolved even by Hinode later.

\section{Vector polarimetry from Hinode}
\label{sec:hinode}

\subsection{Longitudinal fields}

We examined data for quiet and active regions obtained with the 
SP on Hinode.
Such data are
unique, stable, seeing-free measurements of the full Stokes vector,
with pixels of only 0.164\arcsec.  Image stability is critical for
accurate polarimetry since seeing-induced errors can be large
\citep{Lites1987}, even for ground-based images captured with adaptive
optics \citep{Judge+others2004}.  The noise in longitudinal
magnetograms from a ``normal map'' (4.8 second acquisition time)
acquired with the SP corresponds to just $3$ Mx~cm$^{-2}$ in each
pixel \citep{Lites+others2008}.  These data are sensitive to very
small magnetic flux concentrations, the noise levels are  $4\times
10^{14}$ Mx (Table ~\pref{tab:sens}).

Active region data from December 29 2006 and January 19 2007 were
analyzed with quiet data obtained on November 27
2007.  Here we address the quiet (figure~\pref{fig:spqr}) and December
2006 
(figure~\pref{fig:spar}) 
datasets.  Since
the SP is a slit instrument, the regions  took 135 and 43 minutes to
scan respectively.
The
latter small active region contains a small group of pores
and a single pore of opposite positive polarity.
These pores have associated with them
magnetically disturbed granulation.  Figure~\pref{fig:spqrc} shows
close-up views of these magnetic
concentrations,  
again plotting groups of
representative field lines whose lengths are less than and greater
than 10 Mm, and which have flux densities in each pixel greater than
$6$ Mx~cm$^{-2}$ (i.e. $2\sigma$ above the noise).
When
observed at high magnetic sensitivities and spatial resolution, 
the quiet Sun and active
regions have network-like flux concentrations which are surrounded by
mixed polarity fields. The SP data show significant signal almost
everywhere (there is no obvious ``white noise'' component in the
spatial power spectra).  The SP thus 
has not yet reached a limiting small scale of the photospheric
magnetic structure.  

By binning these Hinode
data to 
the pixel size and
sensitivity of the KPVT, we find that, as expected, the KPVT
misses magnetic flux. But the missing flux amounts to just 25\% of 
the detectable flux within the 
supergranular CIs. Thus, although a significant number of short loops connecting NBs to CIs
are certainly missing from figures~\pref{fig:kpvtvault} and
\pref{fig:kpvtvaultsm}, the missing flux would  not be 
sufficient to account for all observed
L$\alpha$ threads as cool loops. 

NB fields are organized into concentrations of various sizes, so care
must be taken in discussing field line lengths and their relation to 
underlying flux and associated UV emission.  It is convenient to
discuss two groups- those unipolar concentrations in excess of $5\times
10^{18}$ Mx, and those below.  

The VAULT-2 and active region SP data fall into the large flux group, the
quiet Sun SP data into the small flux group. In the large flux
group, the SP data contain no examples of small scale ($\la$
$10^{18}$ Mx) mixed polarities within the cores of the network
concentrations themselves, at the detection limit of $\sim 0.4 \times
10^{15}$ Mx, as is suggested by figure~\pref{fig:dowdy86}. If the cool
loop picture is generally applicable, there is a point where we should
observe them.  But the SP data show that, at a resolution of 
240 km, the typical large concentration is unipolar. Any missing 
mixed polarity fields on scales below 240 km would
yield extremely short loops (lengths $\ell \la 240\pi/2 \approx 365$ km)
which would not even be visible in features like L$\alpha$ 
formed higher above the photosphere (see section \pref{subsec:unresolved}). 
The
SP data therefore confirm 
that short loops from larger concentrations associated with active
network (an example might be that at $x=15$ in the lower panel of
figure~\pref{fig:spqrc}) generally only extend from the {\em
peripheries} of these flux concentrations. The core regions of the
concentrations are almost always connected via longer loops between
different concentrations and not between the concentration and CI
field (note the absence of short loops connected to centers of large
concentrations of flux in the lower panel of figure \pref{fig:spqrc}).
The brightest UV and EUV emission from the TR sits
(statistically) directly over the concentrations, as exemplified by
the VAULT-2 data above.  The cool loop picture therefore fails to account
for the bulk of L$\alpha$ emission over such NBs.

The story is different in the small flux group.  The SP data contain
examples of what are probably mixed polarities {\em within} NBs, as
proposed by \citet{Dowdy+Rabin+Moore1986}. For example, if the
negative polarity ribbon of condensations along $x=5$ are considered
part of the larger positive polarity condensations seen along $x=7.5$
in the upper panel of figure \pref{fig:spqrc}, then this situation
qualifies as a mixed polarity network structure as suggested by Dowdy
et al..  The short loops clustered between and around these
concentrations are likely candidates for cool loops, and may explain
the rosette like structures seen in the TR in the quiet Sun (see the
compilation of images from the SUMER instrument on SOHO
\citealp{Feldman+others2003}).

%These regions lie near
%\coords{80}{43} and at the boundaries of a cell whose center is at
%\coords{60}{12}.  In these regions, the field lines appear to be bent
%away from the regions of strong concentrations, though these are some
%10-15Mm distant.

\subsection{Vector fields}

The SP data consist of the full Stokes vectors of the photospheric Fe {\sc i}
6301.5 and 6302.5 \AA\ lines. 
We can therefore assess how the potential
fields compare to the transverse field properties derived from the Stokes $Q,U$
(linear polarization) measurements.  This is meaningful only for the
areas of  active region data with sufficient signal-to-noise
ratios.
Figure~\pref{fig:azimuths} shows the field azimuths both measured and
computed using the MERLIN\footnote{
{\tt  http://www.hao.ucar.edu/projects/csac/nextgen.php\#merlin}.  See
\citet{Lites+others2007}.}  inversion scheme
for the 29 December 2006 dataset.  (Field azimuths are the
angles measured counter-clockwise from the direction pointing solar west,  of the field vector projected on to
the local solar surface. The azimuths are subject to the well known
180$^\circ$ ambiguity).  There is broad agreement.
Significant departures from potential fields exist 
in both 
active region datasets, some arising simply from the limits 
of using Fourier transforms which assume
that the domain is periodic in $x$ and $y$.  (The ``cusps'' seen near 
$y=47$ in
figure \pref{fig:spar} are such artifacts, and the azimuths close to the
boundaries in Figure~\pref{fig:azimuths}  also reflect this problem).  Others are of solar
origin, caused by electrical currents above the surface $z=0$.  
Examples are the many patches of gray near the center of the lower panel of 
in Figure~\pref{fig:azimuths}. 
These differences are of prime interest for the physics of the atmosphere, but
here we ask simply how our conclusions concerning the
validity of the ``cool loop'' model might be changed.

The lack of short loops in the cores of larger network flux
concentrations appears to be a robust result- opposite polarities
simply do not exist there for any length of time.  The directions and
connections of field lines will depart from the potential
calculations.  We do not speculate on such effects, simply noting the
broad agreement in the measured and computed azimuths.  But it is
difficult to see how the ``comet'' structures could remain aligned as
they are seen in the VAULT-2 data by {\em small}-scale current
systems. Indeed, it is well known that filaments must carry
significant electrical currents to provide support against gravity via
the Lorentz force \citep[e.g.][]{Low1996}.  ``Comet''-like structures
seen in upper chromospheric H$\alpha$ are aligned on scales larger
than supergranulation along filament channels
\citep{Martin+others1994}.  The large scale currents associated with
filament channels may therefore also be responsible for the observed L$\alpha$
``comet'' alignment.  As in the potential field case, it is only if
the ``comets'' are associated with large scale ($>$ supergranules)
coronal structure that this observation makes physical sense.

\section{Discussion and conclusions}

On the basis of cool loop models of the VAULT-2 L$\alpha$ emission
analyzed here, 
\citet{Patsourakos+Gouttebroze+Vourlidas2007} concluded that `` The
reasonable agreement between the models and the observations indicates
that an explanation of the observed fine structure in terms of cool
loops is plausible.''  Their motivation for interpretation of the
threads in terms of such models is that the thread-like structure,
assumed to be aligned with magnetic fields associated with the
chromospheric network, is incompatible with heating via field-aligned
heat conduction down from overlying coronal plasma.  The cool loop
model has emerged as a viable explanation for the anomalous brightness
of TR emission in features formed below $2\times10^5$
K, in spite of some significant physical problems, notably the
tendency for instability of such classes of model
\citep{Cally+Robb1991}. Nevertheless, the work by
\citet{Patsourakos+Gouttebroze+Vourlidas2007} lends support to this
picture.

In contrast, by studying VAULT-2 data in terms of the magnetic structure of the chromospheric 
network, we find:

\begin{itemize}
\item{} The location and orientation of some of the L$\alpha$ threads
  are only rarely compatible with the idea of cool, short, loops
  originating from the NBs and extending into
  neighboring CI regions, but 
\item{} such short loops usually connect the CI to the
  {\em peripheries} of network flux  concentrations with unipolar
  fluxes in excess of, say $5\times 10^{18}$ Mx.
\item{}  The bulk of the L$\alpha$ NB emission, arising from the cores of
  such concentrations, seems to be associated with far longer magnetic
  loops
  which connect to other concentrations.  In this way these
  concentrations seem analogous to plages. 
\item{} In the ``quiet'' region of the VAULT-2 data, the 
longer loops diverge non-radially from their NB concentrations in a
manner reminiscent of  the ``comet'' L$\alpha$ patterns. 
\item{}Hinode SP observations at $0\farcs33$ resolution reveal 
  that short loops can exist in what appear to be NBs,
  provided the concentrations making up the magnetic concentrations
  have small enough areas (fluxes $\la 5\times10^{18}$ Mx). 
  Thus, cool loops emitting L$\alpha$ may indeed 
 be present
  in quiet regions, where they may cluster around small concentrations
  of flux, and have the appearance of clumped rosettes of emission of
  $\la 10$Mm diameter (see the upper panel of Figure~\pref{fig:spqrc}).
\item{} Hinode data in  the neighborhood of active
  regions, while showing mixed polarities, tend to reveal short loops
  only connecting peripheries of the larger flux concentrations 
  found there to the CI regions (see the lower panel of
  Figure~\pref{fig:spqrc}).  This is also the case found for 
  the particular VAULT-2 data analyzed
  here. The bulk of bright network 
  L$\alpha$ emission is difficult to accounted for by such structures.
\item{} Non-potential fields are clearly present in the Hinode data,
  but we have argued that such fields, on the small scales associated
  with the cool loop model, do not affect our overall conclusions.
\end{itemize}
 
Our work suggests that the cool loop picture cannot be universally
valid.  Below we propose a different qualitative picture of the emission from
the cores of these flux concentrations.  

\subsection{The ``comet'' patterns seen in VAULT data}

The aligned ``comet-tail'' patterns of the L$\alpha$
threads initially presented us with a puzzle. Why would such large
scale order be characteristic of a small scale process involving the
formation of physically far smaller loops, whose footpoints might be
controlled by random convective processes?  

The explanation may be that the region termed ``quiet'' by Patsourakos and
colleagues is not really ``quiet''.  There is close by a large 
net flux which imposes a large scale order on potential fields. Two other observations
indicate organization on a large scale: the threads largely point away
from the coronal Fe IX/X emitting loops associated with the active
region in the SE corner of the field of view; the ``quiet'' region in
fact lies between two filament channels- the SE channel and another
lying along the western edge of the VAULT FOV (see
fig.~\pref{fig:context}). 
Free magnetic energy associated with the magnetostatic balance of
filaments, in the form of atmospheric current systems, is surely present
which might  modify the overlying magnetic field from the potential
state we have calculated.  Such currents would not be incompatible
with the large scale order implied by the comets. 
It is only if one considers the structures to be formed from
convection-driven short loops that large scale organization would be
surprising.
Our analysis instead points to an association of long loops with most of
the  L$\alpha$ emission, including the threads modeled previously as
cool loops.  Instead of cool loops, the emission seems to 
arise, perhaps as chromospheric
material is launched as spicules and heated along much longer field lines. 
The difficulty in this picture is to explain why L$\alpha$ emission is
bright and extended over several Mm lengths, given the obvious failure of
field-aligned heat conduction to achieve this. 

\subsection{The role of unresolved magnetic fields}
\label{subsec:unresolved}

We noted above that the SP data show signal on all scales down to the
Lunqvist limit (2 pixels $=0\farcs328$).  The smallest scales of solar
photospheric magnetic fields are as yet unknown.  It is therefore
likely that tiny loop structures are missing from our analysis. 
However, the existence of smaller scale structures does not weaken our
conclusions, for several reasons.

Firstly, to the extent that the magnetic fields are potential, 
structures of  horizontal scale $\ell$ in the photospheric normal magnetic field
will extend only to heights $\approx \ell$ higher into the
atmosphere.  This result is a general property of solutions of 
Laplace's equation (see, e.g. \citealp{Gary1989}).  Now L$\alpha$
radiation cannot emerge from heights less than the height where the
continuum optical depth is unity, which occurs near 0.8Mm because of
opacity and atmospheric stratification (FAL). In FAL's models, the
bulk of the emission arises 
at least 2Mm above the photosphere.
Thus, for any small-scale bipoles observable at the
solar surface, {\em only those with footpoints separated by 0.8Mm or
so can 
contribute to observable L$\alpha$ emission}.  Those
separated by smaller scale lengths  likely return to the
photosphere before reaching 1 Mm heights.  The thermal signatures of
such loops would be primarily visible in lines and continua  less opaque than 
L$\alpha$, influencing L$\alpha$ itself only marginally.

Other reasons arise from the nature of the observed L$\alpha$ threads.
They extend over 5-10Mm lengths, so that if formed within tubes of
magnetic flux, footpoint separations are at least this large.  Threads
of lengths between 2 and 5 Mm would extend high enough to emit
L$\alpha$ but are not seen in the data.  The thread orientations are
collectively organized over scales of several supergranules, and the
threads, if both footpoints are anchored in the photosphere, 
require opposite polarity flux surrounding the flux
concentrations associated with network boundaries.  None of these
observations are consistent with random ``salt and pepper''
distributions of flux associated with supergranules on any scale below
say 5 Mm in length.

For all these reasons we believe that unobservably 
small scale magnetic fields are irrelevant to the problem of
understanding the essential properties of the observed L$\alpha$
line, both in plage and other regions.

\subsection{Speculation on the origin of the bulk of NB emission}

Field-aligned diffusive and flowing models of the type computed by FAL 
may indeed account for the moss L$\alpha$ emission, but as noted in
the introduction, they are incompatible with L$\alpha$ thread
emission.  In the absence of cool loops as a viable proposal of
L$\alpha$ emission except in quiet regions, we can speculate on what
might be the cause of the L$\alpha$ properties seen by VAULT.

Figure \pref{fig:tricolor} reveals that the corona overlying the
observed region is bright  on the SE side, dim elsewhere.
The L$\alpha$ emission from network patches
seems to care little of the intensity of overlying coronal emission. 
If energy for L$\alpha$ radiation arises
directly from coronal plasma, surely there must be some
correlation between L$\alpha$ and coronal brightness?  

Independent evidence suggests that the observed L$\alpha$
structure is correlated with observations of H$\alpha$, whose
morphology is complicated, but whose properties on fine scales are
known to relate to conditions in the overlying corona
\citep{Berger+others1999}. 
\citet{Berger+others1999} found that, on scales of arcseconds and less,
coronal ``moss''
emission at 171 \AA{} is dark where  H$\alpha$ wings are strong,
suggesting that the EUV corona and cooler H$\alpha$ plasma are
separated by a thermal interface which lies parallel to magnetic field
lines
(it is not therefore the ``classical'' thin TR, but
more of a sheath.)  We can speculate that this
interface might by a place where energy from the coronal plasma can be
transferred to hydrogen atoms, by diffusion of neutral atoms into the
hot coronal regions \citep{Pietarila+Judge2004}, via cross-field
conduction which occurs because of proton dynamics \citep{Athay1990},
or because of some as yet undetermined (Rayleigh-Taylor like?)
instability \citep{Gabriel1976}.  

Given this interface, consider the neutral atom diffusion scenario
(Judge 2008, submitted).
Neutral atoms, at an interface with hot corona, experience no Lorentz
force until ionized.  The probability that a hydrogen atom is ionized
by collisions with hot coronal particles (electrons) is related to the
probability for excitation to the $n=2$ level, which is almost immediately
($10^{-8}$ s)
followed by the emission of a L$\alpha$ photon. Roughly one L$\alpha$
photon is emitted before it is ionized, independent of the density of
the coronal plasma, provided it is ionized somewhere in the coronal
plasma.  It turns out that the L$\alpha$ intensities expected from
this process are proportional to the thermal energy density of the
corona and the neutral diffusion speed. Thus, there is a different
dependence of coronal and L$\alpha$ emission on thermal parameters, so
it is possible in principle to explain why similar L$\alpha$
intensities may arise from regions with different coronal
intensities. 
The essential difference between this and the cool loop
picture is that coronal thermal energy is drained by diffusion of
neutrals across magnetic field lines to generate much of the emission
seen in L$\alpha$ and 
other typical ``lower TR transitions''.  Some chromospheric process
is assumed to launch spicules to get the process started.  Such a
model can explain the puzzle of the ``comet'' asymmetry noted in
the present paper, in that cool plasma threads along long coronal loops 
already have the large-scale organization required by observations of  the L$\alpha$
threads.  If it proves feasible, this process bypasses thermal
stability problems presented by cool-loop models, and it is appealing
in that the downward directed conductive flux density of $\approx
10^6$ \flxu{}, unaccounted for in cool loop models, is radiated by
cool, strong TR lines.  Further work on this
scenario is in progress (Judge, 2008, submitted).

We thank
Scott McIntosh and the referee for helpful comments on this paper. 
Hinode is a Japanese mission developed and launched by ISAS/JAXA,
collaborating with NAOJ as a domestic partner, NASA and STFC (UK) as
international partners. Scientific operation of the Hinode mission is
conducted by the Hinode science team organized at ISAS/JAXA. This team
mainly consists of scientists from institutes in the partner
countries. Support for the post-launch operation is provided by JAXA
and NAOJ (Japan), STFC (U.K.), NASA, ESA, and NSC (Norway). 

\def\aspcs{{ASP Conf.\ Ser.}}

\clearpage
\tabone
\clearpage
\figone
\figtwo
%\figthree
\figfour
\figfive
\figfivec
\figsix
\figseven
\figeight
\fignine
\end{document}